# Photocatalytic acetaldehyde oxidation in air using spacious TiO$_2$ films prepared by atomic layer deposition on supported carbonaceous sacrificial templates

*Sammy W. Verbruggen,*[1,2,‡] *Shaoren Deng,*[3,‡] *Mert Kurttepeli,*[4] *Daire J. Cott,*[5] *Philippe M. Vereecken,*[2,5] *Sara Bals,*[4] *Johan A. Martens,*[2] *Christophe Detavernier,*[3]* *and Silvia Lenaerts*[1]*

[1] Department of Bio-Engineering Sciences, Sustainable Energy and Air Purification, University of Antwerp, Groenenborgerlaan 171, B-2020 Antwerp, Belgium

[2] Department of Microbial and Molecular Systems, Center for Surface Chemistry and Catalysis, KU Leuven, Kasteelpark Arenberg 23, B-3001 Heverlee, Belgium

[3] Department of Solid State Science, University Ghent Krijgslaan 281 S1, B-9000 Gent, Belgium

[4] Department of Physics, Electron Microscopy for Materials Science (EMAT), University of Antwerp, Groenenborgerlaan 171, B-2020 Antwerp, Belgium

[5] IMEC, Kapeldreef 75, B-3001 Leuven, Belgium

* E-mail: silvia.lenaerts@uantwerp.be; christophe.detavernier@ugent.be

‡These authors contributed equally.

**Abstract**

Supported carbon nanosheets and carbon nanotubes served as sacrificial templates for preparing spacious $TiO_2$ photocatalytic thin films. Amorphous $TiO_2$ was deposited conformally on the carbonaceous template material by atomic layer deposition (ALD). Upon calcination at 550°C, the carbon template was oxidatively removed and the as-deposited continuous amorphous $TiO_2$ layers transformed into interlinked anatase nanoparticles with an overall morphology commensurate to the original template structure. The effect of type of template, number of ALD cycles and gas residence time of pollutant on the photocatalytic activity, as well as the stability of the photocatalytic performance of these thin films was investigated. The $TiO_2$ films exhibited excellent photocatalytic activity towards photocatalytic degradation of acetaldehyde in air as a model reaction for photocatalytic indoor air pollution abatement. Optimized films outperformed a reference film of commercial PC500.

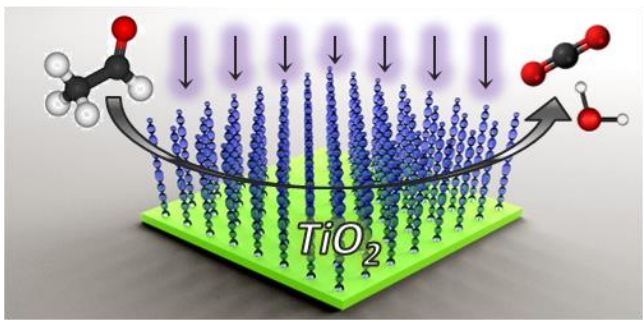

# 1. Introduction

Today's society suffers from the consequences of poor indoor air quality [1]. In this context, photocatalysis is a technology of great interest for air pollution abatement [2,3]. $TiO_2$ is the most encountered photocatalyst and capable of degrading a wide range of contaminants ranging from volatile organic compounds over exhaust gasses to particulate matter and soot [4-8]. The use of commercially available or lab-made photocatalytic powders for gas phase applications is not always evident since immobilization of the nanosized powder particles is vital for safeguarding human health. Direct synthesis of immobilized thin films would mean a big step forward in the development of sustainable photocatalysts for applications involving flowing gas streams. Such films generally have to meet three important design criteria:

> *1. A spacious structure.* Porous catalysts allow gas molecules to easily enter the structure and adsorb on the active sites [9]. The introduction of porosity into a structure using sacrificial templates is a well known strategy [10].

> *2. Controllable film thickness.* Increasing the film thickness will eventually lead to light impermeability, high internal mass transfer resistance and consequently lower efficiency [11].

> *3. Crystallinity and crystallite size.* Most deposition methods lead to uneven films that contain agglomerates of variable size, resulting in a broad particle size distribution. In photocatalysis small, crystalline particles with short charge carrier diffusion distances are commonly favored over large particles that have a higher tendency to generate defects [12,13].

In this work we report on the photocatalytic activity of spacious $TiO_2$ thin films prepared using a highly controllable synthesis strategy. Multiwall carbon nanotubes (CNT in short for this work) and carbon nanosheets (CNS) grown -and thus immobilized- on a silicon wafer were used as templates for $TiO_2$ deposition. By means of atomic layer deposition (ALD) the carbonaceous substrates were coated with a very homogeneous, conformal $TiO_2$ layer. ALD is considered as a deposition method of great potential for producing thin, conformal coatings with thickness control up to the atomic level [14-16]. This results from the specific properties of the ALD process: film growth is achieved by repeating a four-step cycle of self-saturating reactions. It has been demonstrated that ALD is an excellent technique for applying conformal coatings on structures with high-aspect ratios such as immobilized CNTs in this case [17,18]. Alternative approaches that have been applied for coating CNTs such as the sol-gel method [19,20], often lead to rough and thick $TiO_2$ layers. In our samples, transformation of as-deposited amorphous $TiO_2$ into photoactive anatase by calcination coincided with burning the underlying sacrificial carbon template [21,22]. Simultaneously a remarkable morphology change of the $TiO_2$ layer took place. Upon annealing, the as-deposited dense $TiO_2$ layer transformed into cross-linked $TiO_2$ nanoparticles with an overall morphology commensurate to the original underlying template. It is important to note that this morphology is totally different from hollow tube morphology obtained through chemical etching of anodized alumina templates after ALD deposition [23,24]. The material characteristics of $TiO_2$ films obtained by ALD on CNT templates have been documented in detail in our recent work [25]. In the present work the photocatalytic activity in relation to the type of template used (CNT or CNS), the number of ALD cycles and the gas residence time, as well as the stability of the photocatalytic performance of the films are investigated. The combination of techniques in our study led to thin, porous and completely immobilized, high surface area $TiO_2$ nanoparticle films that perform well in gas phase photocatalytic processes such as indoor air purification. Incorporation of the films in heating, ventilation and air conditioning (HVAC) systems could be one of the future applications [26]. A comprehensive characterization of all different catalysts provides more insight in the structure-activity relation of these promising materials.

## 2. Experimental

### 2.1 Thin film synthesis

Carbon nanosheets (CNS) were grown following a recently outlined procedure [27]. In brief, 200 mm diameter Silicon wafers (p-type) were cleaned in a SC1 (APM) mixture to remove any particles. A 100 nm thermal $SiO_2$ layer was grown on the Si wafer followed by a 70 nm TiN layer sputtered from a Ti target in a $N_2$ atmosphere (Applied Endura Extensa TTN). Prior to CNS growth, a $H_2$ plasma pretreatment (300 W) was carried out for 15 min at 0.45 Torr and 750 °C in a capacitively coupled (CC) PECVD reactor with a 13.56 MHz RF generator (Oxford Instruments plasma technology UK. NANOCVD). Then $C_2H_2/H_2$ (1:10) was flown into the chamber and a 300 W plasma at a total pressure of 0.45 Torr was maintained for 30 minutes. The substrate was removed from the chamber and allowed to cool under vacuum (1 x $10^{-4}$ Torr) for 5 min.

For the fabrication of supported multiwall carbon nanotubes (CNTs) a 70 nm TiN layer was deposited similar as above, onto which a 1 nm (nominal) Co layer was sputtered (Applied Endura Extensa TTN). In this case, TiN acts a diffusion barrier to avoid silicidation of the Co catalyst layer. CNTs were grown in a microwave (2.45 GHz) plasma enhanced chemical vapor deposition chamber (PECVD, TEL, Japan). In a typical experiment the Co catalyst layer was exposed to a NH$_3$ plasma for 5 min to transform the film into active metal nanoparticles for CNT growth. Then a C$_2$H$_4$/H$_2$ mixture was flown into the chamber at a temperature of 550°C for 30 min [28].

For the Atomic Layer Deposition process a 5 cm by 1.5 cm piece of grown CNTs or CNS was loaded into a homemade ALD tool with a base pressure in the low 10$^{-7}$ mbar range, as described in earlier work [29]. The sample was placed onto a heated chuck, and heated to 100°C. Tetrakis (dimethylamido) titanium (TDMAT) (99.999% Sigma-Aldrich) and O$_3$ generated by an ozone generator (Yanco Industries LTD) were alternatingly pulsed into the ALD chamber at pressures of 0.3 and 0.5 mbar, respectively. The concentration of ozone in the flux was 145 μg mL$^{-1}$. The pulse (20 seconds) and pump times (40 seconds) were optimized to allow for a uniform coating of TiO$_2$ along the entire carbon template structure and to prevent the occurrence of chemical vapor deposition type reactions. For each type of template, 50, 100, 200 and 400 ALD cycles were applied. After ALD, the samples were calcined at a heating rate of 1°C min$^{-1}$ and kept at 550°C for 3 hours in order to burn off the carbon template and transform as deposited non-crystalline TiO$_2$ into the anatase crystal structure. It is important to note that for convenience purposes we will still refer to the calcined films as 'CNT' or 'CNS' samples, keeping in mind that the original CNT or CNS template has been sacrificially removed during calcination.

A reference sample was prepared by spin coating a Si wafer of the same total dimensions as the ALD samples (5 cm by 1.5 cm) with a layer of PC500 TiO$_2$ powder (Cristal Global, 350 m²g$^{-1}$). This catalyst was selected as a reference because of its anatase nanocrystalline nature and high surface area, displaying similar characteristics as the calcined CNT and CNS samples under study. Furthermore, we have recently shown that this particular photocatalyst is more efficient and cost-effective than the classic P25 for the photocatalytic degradation of acetaldehyde in air [30,31]. The coating was achieved by preparing a suspension containing 150 mg $TiO_2$ in 2 g denaturated ethanol (Royal Nedalco, >99.6%) that was ultrasonically stirred for one hour. Two wafers of 2.5 cm by 1.5 cm were spin coated with one layer of the suspension at 1500 rpm for 1 min. The wafers were dried overnight in an oven at 100°C.

**2.2 Photocatalytic test**

The catalyst films were tested in a custom-made flat bed single pass continuous flow reactor, consisting of a slit-shaped reactor volume of dimensions 150 mm x 20 mm x 2.75 mm. A schematic of the entire reactor set-up is given in Figure 1. The reactor is sealed from the top with a quartz plate. The sample was placed in the middle of the reactor and illuminated with a fluorescent UVA lamp (Philips Cleo, 25 W) placed longitudinally above the long side of the reactor at a distance of 2 cm from the film surface. The light intensity incident on the films was 2.6 mW cm$^{-2}$, as measured with an Avantes AvaSpec-3648 spectrometer.

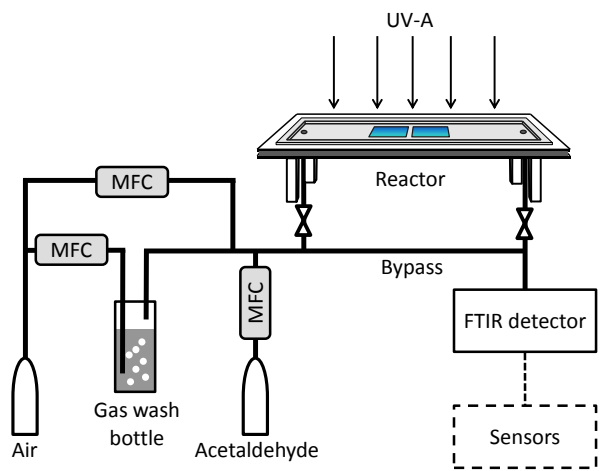

**Figure 1.** Schematic of the photocatalytic test set-up. MFC: Mass Flow Controller.

For all tests acetaldehyde was selected as model compound for indoor air pollution. In a typical experiment air (Air Liquide Alphagaz) spiked with 50 ppmv acetaldehyde (Air Liquide, 1% in $N_2$) at a total flow rate of 400 cm³ min$^{-1}$ is introduced from the bottom on one end of the reactor slit and exits through the bottom at the other end. The relative humidity of the gas stream is kept at 5% (verified using a Vaisala relative humidity sensor) by sending part of the dry airflow through a gas wash bottle before mixing with acetaldehyde. A polluted continuous airflow is sent through the reactor under dark conditions for 15 min in order to establish adsorption equilibrium. Consecutively the UVA lamp is switched on for 20 min. For all samples a continuous steady state acetaldehyde degradation level was reached within the first 5 min of illumination. Blanc tests with an empty reactor and a reactor loaded with uncoated silicon wafers showed no acetaldehyde mineralization under UVA illumination. The temperature in the reactor was measured using a thermocouple and did not exceed 30°C.

Continuous steady state acetaldehyde to $CO_2$ mineralization is studied by on-line FTIR gas phase spectroscopy of the reactor outlet stream, as described in our previous work [30-34]. In short, the acetaldehyde and $CO_2$ concentrations were monitored in time using the MacrosBasic software (Thermo Fisher) by recording the change in FTIR peak heights of the acetaldehyde $\upsilon_{H-C=O}$ stretching vibration at 2728 cm$^{-1}$ and the $\upsilon_{C=O}$ stretching vibration of $CO_2$ at 2360 cm$^{-1}$. These wave numbers were carefully selected, as they do not interfere with other bands in the FTIR spectrum. The FTIR absorbance related to both species is in turn converted into actual concentrations by means of calibration curves constructed using a Dräger Polytron Organic Vapor sensor and a Vaisala $CO_2$ sensor. The acetaldehyde inlet concentration is determined from the steady FTIR peak height level of the $\upsilon_{H-C=O}$ stretching vibration at 2728 cm$^{-1}$ at the end of the adsorption equilibrium phase, more specifically during the final five minutes of the dark phase. The mineralization is determined as the steady state $CO_2$ formation level, calculated from the IR peak height during the final ten minutes of the UVA illumination phase. As a final comment it is good to note that the acetaldehyde concentrations used in our tests are significantly higher than regular indoor air exposure levels. This is due to the detection limit of ca. 0.7 ppmv of the FTIR apparatus.

## 2.3 Instrumentation

After cleaving the sample, cross sections of the $TiO_2$ coated CNS and CNT templates were characterized by scanning electron microscopy (FEI Helios NanoLab 650 dual-beam system). For TEM imaging, the sample surface was first scraped off into an agate mortar and then a suspension was made by diluting the scrapings in ethanol. A droplet of this suspension was deposited onto a carbon-coated holey film on a Cu grid. TEM experiments were carried out on a Philips CM30-FEG microscope operated at 300 kV. X-ray fluorescence measurement was performed in a Bruker Artax system including a Mo X-ray source and an XFlash 5010 silicon drift detector. X-ray diffraction spectra were collected with a Bruker D8 system in the range of 20° - 60° 2θ. UV-VIS thin film absorbance spectra were collected in the range of 300-700 nm using a Shimadzu UV-2501PC double beam spectrophotometer equipped with a 60 mm $BaSO_4$ coated integrating sphere and a Photomultiplier R-446U detector. A custom-made sample holder was used to fixate the wafers in order to probe the same area in the center of the film for every sample. To calculate the surface area (enhancement) of the porous $TiO_2$ films, ex situ XRF characterization was employed as discussed in previous work [35]. Basically, 50 cycles of ZnO were deposited onto the $TiO_2$ nanostructured samples, as well as on a control sample (consisting of a planar Si substrate after the same ALD $TiO_2$ deposition and subsequent annealing process) by thermal ALD using Diethylzinc (Sigmal Aldrich) and $O_3$ at 100°C. To achieve conformal coating and avoid chemical vapor deposition in such porous structure, 40 s pulse time and 60 s pumping time were applied. Afterwards, by comparing the Zn XRF signal intensities between $TiO_2$ nanostructured samples on one hand and control sample on the other hand, the surface area increase induced by the $TiO_2$ nanostructure compared to a planar substrate can be deduced. For the photocatalytic tests, the gaseous species in the reactor outlet were detected using on-line FTIR spectroscopy. The apparatus was a NicoletTM 380 (Thermo Fisher Scientific) equipped with ZnSe windows and a 2 m heated gas cell. Five spectra were collected every minute in the wavenumber range 4000-400 $cm^{-1}$ at a resolution of 1 $cm^{-1}$.

## 3. Results and discussion

### 3.1 Catalyst Characterization

Photographs of CNT and CNS template based films after $TiO_2$ ALD and calcination at 550°C are shown in Figures 2a and 2b respectively. Deposition of 50 ALD cycles on CNS results in a film that does not look much different from the underlying silicon wafer. As more $TiO_2$ is deposited, the films become shiny with a grey-blue-violet sheen. The samples prepared using the 10 μm thick CNT template appear shiny blue after 50 and 100 ALD cycles, but become dull and grayish-white after 200 and 400 ALD cycles. The spin coated PC500 reference film results in a $TiO_2$ layer with an average thickness of ca. 1 μm (Figure 2c). Figures 2d-f show electron microscopy images of the 200 ALD on CNS sample. It can be observed that calcination of the film did not lead to destruction of the sheet structure and the overall original template morphology was retained. The image of the as-deposited sample is in good agreement with similar coated structures reported in literature [36,37]. The TEM detail demonstrates that after calcination the $TiO_2$ layers are now arranged as a network of crystalline nanoparticles. The same pictures are also on display for the 100 ALD on CNT sample (Figures 2g-i). Also in this case, calcination did not lead to destruction of the structure, as the vertical alignment of CNTs is still clearly recognizable. This indicates that the entire tube bed was uniformly coated with TiO$_2$ by ALD. If not, collapse of the structure is to be expected. The TEM detail confirms that the calcination step resulted in the transformation of continuous TiO$_2$ layers as deposited by ALD, into chains of crystalline TiO$_2$ nanoparticles. The precise control of the TiO$_2$ layer thickness offered by ALD, together with the open structure offered by the carbonaceous templates, clearly fulfill the first two design criteria discussed in the introduction.

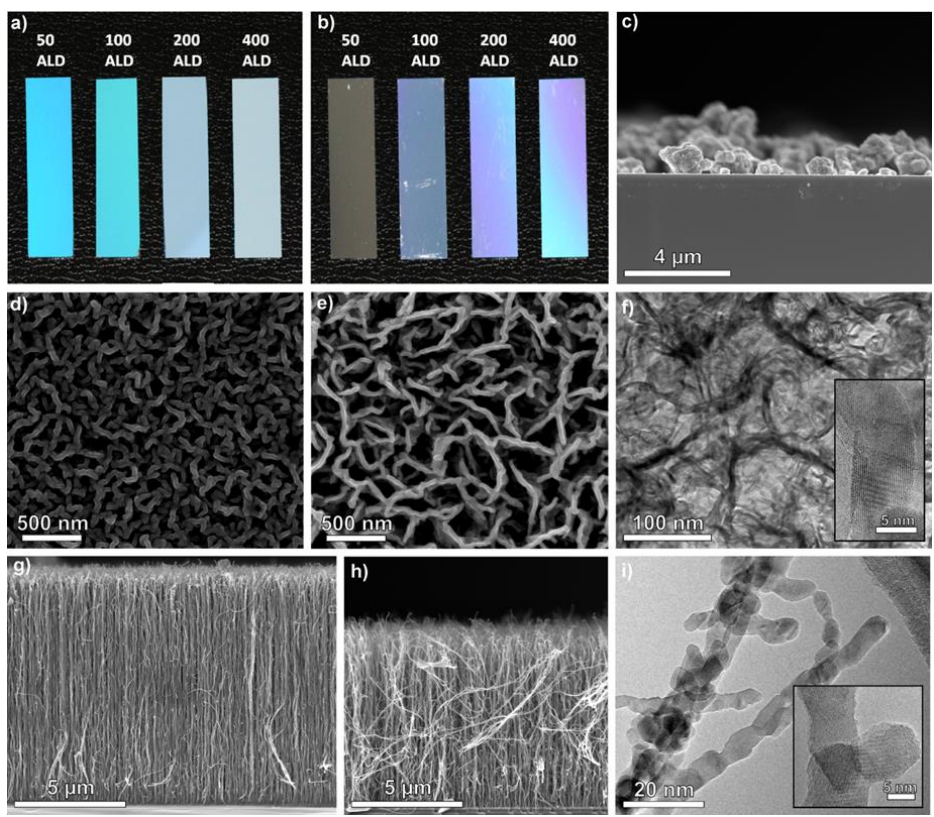

**Figure 2.** Picture of 50-400 TiO2 ALD cycles on 10 μm CNT template after annealing (a) and on CNS template after annealing (b), cross section SEM image of PC500 reference film (c), SEM image of 200 TiO2 ALD cycles on CNS template as deposited (d), after calcination (e) and TEM details of the latter (f). Cross section SEM image of 100 TiO2 ALD cycles on CNT template as deposited (g), after calcination (h) and TEM details of the latter (i).

XRD spectra of both types of calcined films in Figure 3 show the presence of $TiO_2$ in the anatase crystal structure for CNT as well as CNS sacrificial template samples after ALD and calcination at 550°C, also fulfilling the third design criterion. The XRD spectrum of the CNT sample also shows a minor feature at 36.1° 2θ attributed to the (101) plane of rutile that may be formed after partial oxidation of the TiN layer underneath the tubes.

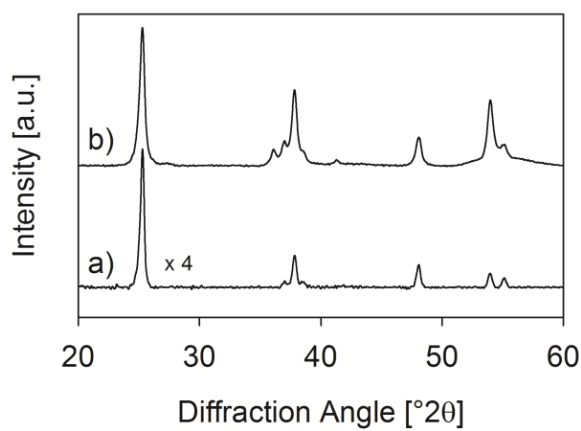

**Figure 3.** XRD pattern of a) 200 $TiO_2$ ALD cycles on CNS and b) 100 $TiO_2$ ALD cycles on CNT, both calcined at 550°C.

The relative $TiO_2$ content determined by XRF and relative UV absorbance of the ALD treated samples with respect to the PC500 reference sample are given in Table 1. Concerning the CNS samples, 50 and 100 cycles lead to a lower $TiO_2$ amount deposited on the wafers compared to the PC500 reference, whereas 200 and 400 cycles lead to a higher $TiO_2$ content. The increase in $TiO_2$ with increasing number of ALD cycles is linear up to 200 ALD cycles, indicating that the pores in the CNS template are filled completely when depositing over 200 ALD cycles of $TiO_2$. It also becomes clear that only the sample containing 400 ALD cycles on CNS is capable of absorbing more UV light compared to the reference and the UV absorbance of the samples increases with increasing TiO$_2$ content. For the CNT template samples, even 50 ALD cycles already lead to a TiO$_2$ content that is a factor six higher compared to the PC500 reference. This is due to the high surface area available for ALD provided by the 10 μm thick CNT forest. The TiO$_2$ content further increases drastically with increasing number of ALD cycles. Also in this case, the TiO$_2$ content increases linearly with the number of ALD cycles up to 200 cycles, again indicating that from that moment on the interstitial voids between the tubes are being filled entirely. For all samples, the UV absorbance is higher than that of the reference. Although the trend is not very clear, roughly, a decrease in UV absorbance with increasing TiO$_2$ content can be observed, which can be attributed to the increasing contribution of scattering from the films, which is in turn intuitively evidenced by the greyish-white appearance at higher TiO$_2$ content (Figure 2a).

**Table 1.** Relative TiO$_2$ content and UV absorbance of TiO$_2$ thin films prepared by different numbers of ALD cycles on CNS or CNT templates with regard to the PC500 reference coating

|  | Number of TiO$_2$ ALD cycles | Relative TiO$_2$ content [a] | Relative UV absorbance [b] |
|---|---|---|---|
| PC500 (reference) | *N.A.* | 1 | 1 |
| CNS template | 50 | 0.2 | 0.3 |
|  | 100 | 0.7 | 0.6 |
|  | 200 | 1.7 | 0.9 |
|  | 400 | 2.2 | 1.1 |
| 10 µm CNT template | 50 | 6.7 | 1.9 |
|  | 100 | 14.9 | 1.5 |
|  | 200 | 28.8 | 1.3 |
|  | 400 | 36.3 | 1.5 |

[a] Determined by the ratio of the Ti XRF signal of the film over the Ti XRF signal of the PC500 reference coating. Due to the lower XRF accuracy at thicker TiO$_2$ layers, a slight underestimation of the content is possible (max. 5-10%).

[b] Determined as the ratio of the integrated UV absorbance between 320 nm and 390 nm of the UV-VIS thin film absorption spectrum, over the integrated UV absorbance of the PC500 reference coating in the same wavelength region.

By comparing the linear increase in TiO$_2$ content with increasing number of ALD cycles for a given carbonaceous template, with the linear TiO$_2$ content increase with increasing number of ALD cycles on a flat silicon wafer, conclusions can be drawn concerning the surface area enhancement induced by the template [35]. For the CNS template, this method reveals a surface area enhancement by a factor of 13, whereas for the CNT template the surface area is enhanced by a factor of 235. Stated otherwise, every cm² of silicon wafer contains 13 cm² or 235 cm² of TiO$_2$ surface for the as-deposited films (prior to calcination) on the CNS and CNT templates respectively. As mentioned above, this is only valid up to 200 ALD cycles, as a higher number of cycles leads to sealing of the pores and causes less sites to be accessible. After calcination at 550°C, the surface area enhancement factors were determined in a similar way and amounted to 2.3 and 260 for the CNS and CNT sacrificial templates, respectively. For 400 ALD cycles on CNT, the surface area enhancement after calcination was determined to be only 175, again indicating that interstitial voids are being filled at this point and the structure has become denser and thus less accessible. The low surface area enhancement for the CNS samples after annealing is attributed to shrinkage of the film thickness (from ca. 500 to 400 nm for the 200 ALD on CNS sample) and densification of the structure by sintering. These phenomena can be explained by the observations of Rooth et al., who noticed the presence of uncoated graphene layers at the bottom of the CNS film [36]. These will induce instability by their removal upon annealing, leading to collapse in the bottom part of the film. On the other hand they also observed the presence of pinholes in the $TiO_2$ on CNS coating. During calcination and removal of the CNS substrate, these pinholes present weak spots in the structure where different $TiO_2$ layers will crumble together and sinter into a rigid structure with loss of surface area as a consequence.

As a final remark, it should be mentioned that we have tried to include a PC500 reference sample with a $TiO_2$ content comparable to that of the CNT samples. This was, however, not possible due to the poor adhesion of the required amount of PC500 on this size of silicon wafer.

**3.2 Photocatalytic activity measurements**

The photocatalytic activities of all films, represented by the steady state acetaldehyde to $CO_2$ mineralization rate are depicted in Figure 4. The acetaldehyde inlet concentration was 50 ppmv, hence the theoretical limit of $CO_2$ formation is 100 ppmv. The PC500 reference film leads to a continuous degradation of 19.5 ppmv min$^{-1}$ of acetaldehyde (a conversion of almost 40%), corresponding to the formation of 39 ppmv min$^{-1}$ of $CO_2$. Concerning the films prepared using the CNS sacrificial template, only the sample with 200 cycles performed better than the reference. This can be attributed to the 1.7 times higher $TiO_2$ content, although the UV absorbance is slightly lower (Table 1). Nevertheless, it is important to note that from Table 1 and Figure 4 it can be deduced that a calcined CNS sample with 100 to 200 $TiO_2$ ALD cycles deposited and subsequent annealing at 550°C, results in a very photocatalytically active, conformal, thin $TiO_2$ film, comparable to the very efficient PC500 catalyst. The occurrence of a maximum in photocatalytic activity for the 200 ALD cycle samples can be well rationalized. At low deposited $TiO_2$ amounts, increasing the $TiO_2$ content leads to more active sites and higher UV absorbance, i.e. more use of incident photons for generating charge carriers. As discussed in the previous section, exceeding 200 ALD cycles results in sealing of the film voids, rendering less active sites to be available for reaction.

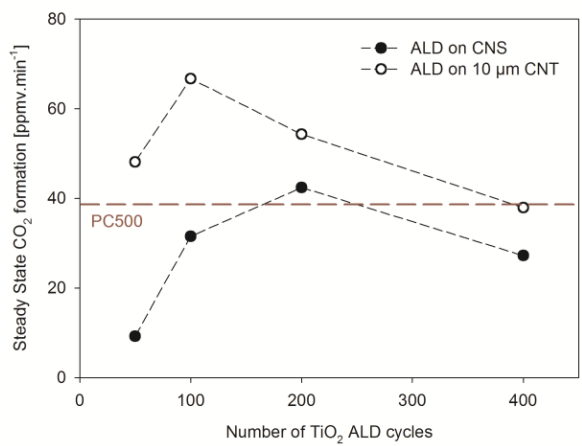

**Figure 4.** Steady state $CO_2$ formation in function of the number of ALD cycles on CNS (filled symbols) or CNT (open symbols) sacrificial templates by the photocatalytic mineralization of a continuous flow of 50 ppmv min$^{-1}$ acetaldehyde in air. The red dashed line represents the activity of the PC500 reference (irrespective of the number of ALD cycles).

For the films prepared on the CNT sacrificial templates, a similar maximum is observed for the sample with 100 ALD cycles. All CNT template samples perform at least as good as the PC500 reference. This is not surprising taking into account the much higher $TiO_2$ content, higher UV absorbance and the high surface area enhancement offered by the 10 μm thick CNT template. Still, it is important to note that the improvement in activity for the best CNT sample compared to PC500 (factor 1.7) is far less than the increased $TiO_2$ content (factor 14.9). On the other hand, the increase in UV absorbance (factor 1.5) is of the same order of magnitude. The same applies to the difference in activity between CNT samples compared to CNS samples.

### 3.3 Variation gas residence time over catalyst

For this test, the acetaldehyde concentration was kept constant at (23 ± 1) ppmv in air, whereas the total flow rate was varied between 400 and 800 cm³ min$^{-1}$. A higher the activity at shorter residence times represents a more efficient catalyst film. The results are shown in Figure 5. The first important observation is that the reactivity order of all catalysts is the same as in Figure 4. In total, five out of eight samples are more efficient than PC500 (all ALD on CNT samples and the 200 cycles on CNS sample). Moreover, the best performing sample (100 ALD cycles on CNT template) closely approaches the stoichiometric limit of 46 ppmv min$^{-1}$ steady state $CO_2$ formation, making this film almost twice as active as the PC500 commercial reference at short residence times (0.15 s). What is also interesting to note, is that the best performing CNS sample (200 ALD cycles) outperforms the PC500 reference and even the 400 ALD on CNT sample, despite its lower $TiO_2$ content and UV absorbance. This can again be explained by the fact that the structure of the 200 ALD on CNS sample is more open and accessible than the 400 ALD on CNT sample, which is denser because its excessive $TiO_2$ loading.

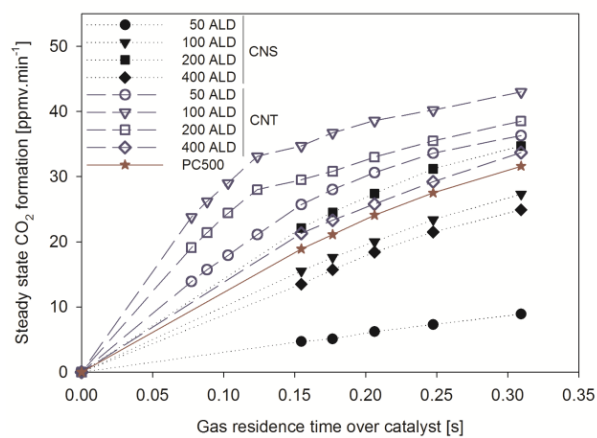

**Figure 5.** Steady state $CO_2$ formation rate in function of the gas residence time over the catalyst films for a constant acetaldehyde inlet concentration of ca. 23 ppmv for the catalyst samples prepared by ALD on CNS (filled symbols, dotted black lines), ALD on 10 μm CNTs (open symbols, dashed blue lines) and the PC500 reference (star symbol, solid red line).

### 3.4 Performance stability

In order to verify the stability of the prepared catalyst films, the best performing sample (100 ALD cycles on CNT) was subjected to ten consecutive photocatalytic test runs (Figure 6). The sample was not subjected to any intermediate washing or cleaning steps and thus repeatedly underwent reactor-flushing, dark-adsorption and UV-degradation phases (same conditions as the experiments in Figure 4). As can be observed very clearly, the photocatalyst film maintains its activity, even after ten consecutive uses.

We conclude that ALD deposition of $TiO_2$ on a sacrificial carbonaceous templates offers an attractive synthesis route for creating well immobilized, efficient, spacious and stable thin film catalysts for photocatalytic acetaldehyde degradation, which can be considered as a model reaction for indoor air purification.

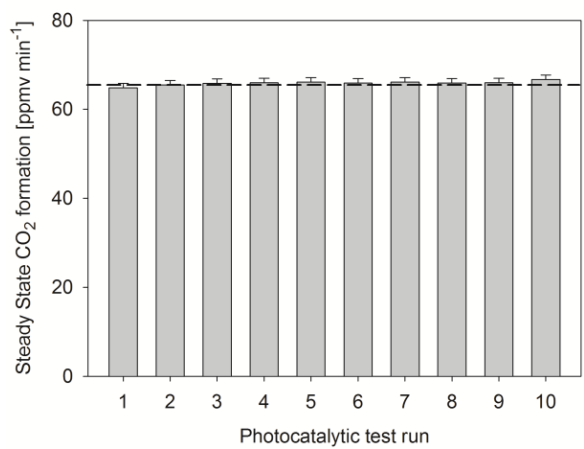

**Figure 6.** Steady state acetaldehyde to $CO_2$ mineralization rate over the $TiO_2$ film prepared by 100 ALD cycles on CNT sacrificial template during each of ten consecutive photocatalytic test runs.

## 4. Conclusions

Carbon nanosheets and carbon nanotubes are attractive templates for preparing photoactive $TiO_2$ films by means of atomic layer deposition. After calcination of the films at 550°C, the template was removed and the as-deposited continuous amorphous $TiO_2$ layers transformed into a network of crystalline $TiO_2$ anatase nanoparticles, commensurate to the overall template morphology. For both types of carbonaceous templates an initial increase in photocatalytic activity was observed with increasing amounts of deposited $TiO_2$. Excessive amounts of $TiO_2$ led to less open films and consequently lower photocatalytic activities. Application of 100 ALD cycles led to the optimum amount of $TiO_2$ deposited on CNTs, whereas the optimum for the CNS templates was reached after 200 ALD cycles. All together, the 100 ALD on CNT sample showed the highest activity. It was almost twice as efficient as the PC500 reference film at a gas residence time as short as 0.15 s and its high photocatalytic activity was maintained even after ten consecutive test runs. In conclusion, we have demonstrated that CNT and CNS sacrificial templates in combination with ALD can be used for the controlled deposition of well-immobilized, thin, conformal, spacious and very active $TiO_2$ films that are excellent for use in gas phase photocatalytic degradation reactions.

**Acknowledgements**

S.W.V. wishes to thank the Research Foundation - Flanders (FWO) for the financial support. C.D. and S.B. acknowledge the European Research Council for funding under the European Union's Seventh Framework Programme (FP7/2007-2013) / ERC grant agreement n° 239865-COCOON and n° 335078-COLOURATOM. J.A.M. acknowledges the Flemish government for long-term structural funding (Methusalem). J.A.M. and S.B. are grateful to the federal government for support in the IAP-PAI P7/05 project.